\documentclass[11pt,cite]{article}
\usepackage{latexsym}

\usepackage[latin1]{inputenc}

\usepackage{graphicx}
\usepackage{color,pst-plot}

\usepackage{amsmath}
\usepackage {amsfonts,amssymb,amstext}

\newcommand{\lbl}[1]{\label{eq:#1}}
\newcommand{ \rf}[1]{(\ref{eq:#1})}

\newcommand{\be}{\begin{equation}}
\newcommand{\ee}{\end{equation}}
\newcommand{\bea}{\begin{eqnarray}}
\newcommand{\eea}{\end{eqnarray}}
\newcommand{\setl}{\setlength\arraycolsep{2pt}}

\newcommand{\noi}{\noindent}
\newcommand{\nn}{\nonumber}
\newcommand{\ra}{\rightarrow}

\newcommand{\cA}{{\cal A}}

\newcommand{\cC}{{\cal C}}

\newcommand{\cF}{{\cal F}}

\newcommand{\cL}{{\cal L}}
\newcommand{\cM}{{\cal M}}

\newcommand{\cO}{{\cal O}}

\newcommand{\Imm}{\mbox{\rm Im}}

\newcommand{\tr}{\mbox{\rm tr}}
\newcommand{\Li}{\mbox{\rm Li}}
\newcommand{\MeV}{\mbox{\rm MeV}}
\newcommand{\GeV}{\mbox{\rm GeV}}

\newcommand{\annd}{\mbox{\rm and}}
\newcommand{\foor}{\mbox{\rm for}}

\newcommand{\stern}{\langle\bar{\psi}\psi\rangle}

\input epsf

\def\theequation{\arabic{section}.\arabic{equation}}

\begin{document}

\begin{titlepage}

\begin{flushright}
\end{flushright}

\vspace*{0.2cm}
\begin{center}
{\Large {\bf The Constituent Chiral Quark Model Revisited~\footnote{Partially based on my talk at the INT Workshop on ``The hadronic Light--by--Light Contribution to the Muon Anomaly'', February 28th -March 4th, 2011.} }}\\[2 cm]

{\bf Eduardo de Rafael}\\[1cm]

 {\it Centre  de Physique Th{\'e}orique~\footnote{Unit{\'e} Mixte de Recherche (UMR 6207) du CNRS et des Universit{\'e}s Aix Marseille 1, Aix Marseille 2 et Sud Toulon-Var, affili{\'e}e {\`a} la FRUMAM.}\\
       CNRS-Luminy, Case 907\\
    F-13288 Marseille Cedex 9, France}

\end{center}

\vspace*{3.0cm}

\begin{abstract}
We reconsider the Constituent Chiral Quark Model of Manohar and Georgi in the presence of $SU(3)_{L}\times SU(3)_{R}$ external sources. As recently emphasized by Weinberg, the corresponding effective Lagrangian is renormalizable  in the Large--${\rm N_c}$ limit. We show, however, that the number of the required counterterms depends crucially on the value of $g_A$ and it is minimized for $g_A =1$. We then find that with a rather small value for the constituent quark mass, which we fix phenomenologically to $M_Q =(190\pm 40)\MeV$, the model reproduces rather well the values of several well known low energy constants. We also comment on the limitations of the model as well as on a few {\it exceptional} applications, to more complicated low--energy observables, where one can expect the model to make reasonably good predictions.
\end{abstract}

\end{titlepage}

\section{\normalsize Introduction}\lbl{int}
\setcounter{equation}{0}
\def\theequation{\arabic{section}.\arabic{equation}}

The Constituent Chiral Quark Model (C$\chi$QM)~\cite{MG84} emerged as an attempt to reconcile  the  successes of phenomenological quark models, like the De~R\'ujula-Georgi-Glashow model~\cite{deRGG75}, with Quantum Chromodynamics (QCD). The corresponding Lagrangian proposed by Manohar and Georgi (MG) is an effective field theory which incorporates the interactions of the low--lying pseudoscalar particles of the hadronic spectrum, the Nambu-Goldstone modes of the spontaneously broken chiral symmetry (S$\chi$SB), to lowest order in the chiral expansion~\cite{Wei79}, and in the presence of chirally rotated quark fields which, because of the S$\chi$SB, have become massive. These constituent quark fields are assumed to have gluonic interactions as well but, since the Goldstone modes are already in the Lagrangian, the color--${\rm SU(3)}$ coupling constant  is supposed to be no longer running and relatively small. The hope is that such an effective Lagrangian encodes the essential degrees of freedom to describe Hadron Physics at energies below the chiral symmetry breaking scale but above the confinement regime.
 
It is fair to say, however, that in spite of some efforts (see e.g. ref.~\cite{EdeRT90} and references therein), it has not been possible to establish the approximations at which the MG--Lagrangian could be derived from the underlying QCD theory. It can be shown to be a particular case of the Extended Nambu Jona-Lasinio (ENJL) Model~\cite{NJL61,BBdeR93}, but this only translates the problem of its derivation from first principles to yet another level.
 
A question to ask which may perhaps be simpler to answer is what are the approximations at which these low--energy models like the MG--model and the ENJL--model could follow from QCD in the limit of a large number of colours ${\rm N_c}$~\cite{tH74}. This is a  limit where the hadronic spectrum of QCD consists of an \underline{infinite} number of narrow states~\cite{Wi79}. In that respect what has been shown~\cite{PPdeR98} is that, in the ENJL--Model, when the unconfining $Q\bar{Q}$ pairs of constituent quarks which contribute to the physical  spectral functions, and which violate the QCD Large--${\rm N_c}$ counting rules,  are removed by adding an appropriate series of local counterterms, what results is an effective Resonance Chiral Lagrangian  with three narrow states: V(vector), A(axial-vector) and S(scalar) of the type discussd in refs.~\cite{EGPdeR89,EGLPdeR89}. These Resonance Chiral Lagrangians, and their extensions~(see e.g. ref.~\cite{CENP03} and references therein), can then be viewed as  simplified versions of the Large--${\rm N_c}$ QCD Hadronic Lagrangian when limited to a \underline{finite} number of states. Integrating out the heavy V, A and S states results  in specific predictions for the $\cO(p^4)$ couplings of the chiral Lagrangian~\cite{GL85} as well as for the higher order terms. When confronted with the phenomenological determinations of these couplings, the predicted values turn out to be rather good (see e.g. ref.~\cite{BBdeR93}).

In full generality, the couplings of the Effective Chiral Lagrangian of the Strong Interactions can be identified with the  coefficients of  the Taylor expansion of appropriate QCD Green's Functions. By contrast, most of the couplings of the Effective Chiral Lagrangian of the Electroweak  Interactions correspond to integrals over all the range of euclidean momenta of appropriate two--point functions with soft insertions of local operators. Their determination, therefore,  requires a precise matching of the short--distance and the long--distance contributions to the underlying QCD Green's functions, which effective Lagrangians like the MG and the ENJL--models do not provide in general. It is because of this that, in phenomenological applications, they have progressively  been replaced by a more direct approach where the relevant Green's functions are approximated by a \underline{finite} number of the Large--${\rm N_c}$ QCD narrow states. Here, the methodology~(see ref.~\cite{EdeR03} for a review), consists in fixing the couplings and masses of a minimal ansatz of narrow states which approximate a specific Green's function in such a way that, on the one hand the short--distance behaviour predicted by the operator product expansion (OPE)~\cite{SVZ79} of the underlying Green's function in Large--${\rm N_c}$ QCD is satisfied and, on the other hand,  the long--distance behaviour constraints governed by the Effective Chiral Lagrangian of the Strong Interactions are satisfied as well. This approach has led to a remarkable set of interesting predictions for some of the couplings of the Electroweak Lagrangian in the chiral limit~(two representative references are~\cite{PdeR00,HPdeR03}). The incorporation of chiral corrections, however, becomes technically rather cumbersome and, above all, the question of the reliability of the approximation with a finite number of narrow states to Large--${\rm N_c}$ QCD remains open (see e.g. ref.~\cite{MP07}).

Concerning the matching between long and short--distances, we would like to point out that 
there is, however,  a class of  low--energy observables, governed by integrals of  specific QCD Green's functions, for which  the MG--Lagrangian predictions, in spite of its limitations, could be rather reliable. This is the case when the 
leading short--distance behaviour of the underlying Green's functions of a given observable is governed by perturbative QCD. Interesting examples of this class of observables   are the Hadronic Vacuum Polarization and the Hadronic Light--by--Light Scattering contributions to a low--energy observable like the anomalous magnetic moment of the muon: $\frac{1}{2}(g_{\mu}-2)$. Furthermore, as recently pointed out by Weinberg~\cite{Wei10}, the MG--Lagrangian in the Large--${\rm N_c}$ limit,  modulo the addition of a finite number of local counterterms, is a renormalizable Lagrangian. Calculations with the MG--Lagrangian, compared to those with the more sophisticated approaches described above, have the advantage of simplicity and, when applied to this class of low--energy observables, can provide a check to the more elaborated phenomenological approaches. These are the reasons which, in our opinion, justify a reconsideration of the MG--Lagrangian. 

We have organized this paper in the following way. The effective Lagrangian of the C$\chi$QM we propose, which also incorporates couplings to external sources, is described in the next section. The corresponding predictions for the $\cO(p^4)$ couplings in the chiral expansion~\cite{GL85} are discussed in Section III, where we also present a discussion on the phenomenological determination of the constituent quark mass $M_Q$. Section IV is dedicated to the study of the Left--Right Correlation function in the C$\chi$QM where both the resulting good and bad features are discussed. Finally, the discussion of the $\pi^0 \ra e^+ e^-$ decay in the C$\chi$QM is discussed in Section V. We summarize our conclusions in Section VI.

\section{\normalsize The Effective Lagrangian}
\setcounter{equation}{0}
\def\theequation{\arabic{section}.\arabic{equation}}

We propose to consider the following effective Lagrangian:

{\setl
\bea\lbl{CCQL}
\cL_{{\rm C}\chi{\rm QM}}(x) & \!\! =\!\! & \underbrace{i{\bar Q}\gamma^{\mu}\left(\partial_{\mu}+\Gamma_{\mu}+iG_{\mu} \right)Q-\frac{i}{2}{g_A}\ {\bar Q}\gamma^{\mu}\gamma_5 \xi_{\mu}Q-M_{Q} {\bar Q}Q}_{\it M-G}-\frac{1}{2}{\bar Q}\left(\Sigma -\gamma_5 \Delta \ \right)Q \nn\\ 
 & + & \underbrace{\frac{1}{4}{ F_{\pi}}^2 \tr\left[ D_{\mu}UD^{\mu}U^{\dagger}\right.}_{\it M-G}
+\left. U^{\dagger}\chi+\chi^{\dagger}U\right]-\underbrace{\frac{1}{4}\sum_{a=1}^{8} G_{\mu\nu}^{(a)}G^{(a)\mu\nu}}_{\it M-G}+ e^2 { C}\ \tr (Q_R U Q_L U^{\dagger}) \nn \\
 & + & 
{ L_5}\  \tr D_{\mu}U^{\dagger}D^{\mu}U(\chi^{\dagger}U+U^{\dagger}\chi) +
{ L_8}\  \tr (U\chi{\dagger}U\chi{\dagger}+U^{\dagger}\chi U^{\dagger}\chi)\,.
\eea}

\noi
The underbraced terms are  those of the MG--Lagrangian, but in the presence
of external $SU(3)$ vector $v_{\mu}(x)$ and axial-vector $a_{\mu}(x)$ sources. The field matrix $U(x)$ is a  3$\times$3 unitary matrix in flavour space which collects the Goldstone fields and which under chiral $SU(3)_{L}\times SU(3)_{R}$ rotations $(V_{L},V_{R})$ transforms as $U\ra V_{R} U V_{L}$. The vector field matrix $D_{\mu}U$ is the covariant derivative of $U$  with respect to the external sources:
\be
D_{\mu}U=\partial_{\mu}U-ir_{\mu}U+iUl_{\mu}\,,\quad l_{\mu}=v_{\mu}-a_{\mu}\,,\quad r_{\mu}=v_{\mu}+a_{\mu}\,, 
\ee 
and, with $U=\xi\xi$,
\be
\Gamma_{\mu}=\frac{1}{2}\left[
\xi^{\dagger}(\partial_{\mu}-ir_{\mu})\xi+
\xi(\partial_{\mu}-il_{\mu})\xi^{\dagger}\right]\,,\quad
\xi_{\mu}=i\left[\xi^{\dagger}(\partial_{\mu}-ir_{\mu})\xi-
\xi(\partial_{\mu}-il_{\mu})\xi^{\dagger}\right]\,.
\ee
The gluon field matrix in the fundamental representation of color $SU(3)$ is $G_{\mu}(x)$ and $G_{\mu\nu}^{(a)}(x)$ its corresponding gluon field strength tensor.
The presence of external scalar $s(x)$ and pseudoscalar $p(x)$ sources  induces the extra terms proportional to
\be
\chi= 2 B [s(x)+i p(x)]\,,
\ee
where $B$, like $F_{\pi}$, is an order parameter which has to be fixed from experiment. When these sources are frozen to the up, down, and strange light quark masses of the QCD Lagrangian, 
\be
\chi= 2 B \cM\,,\quad {\rm with}\quad \cM={\rm diag}(m_u\,, m_d\,, m_s)\,,
\ee
and then
\be
\Sigma=\xi^{\dagger}\cM \xi^{\dagger}+\xi\cM^{\dagger} \xi\,,\quad 
\Delta=\xi^{\dagger}\cM \xi^{\dagger}-\xi\cM^{\dagger}\xi\,.
\ee

With the axial coupling  fixed to  $g_A =1$,
the extra couplings $L_5$ and $L_8$
are the only terms which are needed to absorb the ultraviolet (UV) divergences  when the constituent quark fields $Q(x)$ are integrated out~\footnote{We disregard divergent couplings involving external fields alone to lowest order in the chiral expansion.}. If one wants to consider the case where photons are also integrated out then, to leading order in the chiral expansion and in the electric charge coupling $e$, the last term in the second line is also required to absorb further UV--divergences. This term will be discussed in detail in Section IV. 
Loops involving pion fields are subleading in the $1/{\rm N_c}$--expansion and hence, following the observation of Weinberg in ref.~\cite{Wei10}, the Lagrangian in Eq.~\rf{CCQL}, when considered within the framework of the large--${\rm N_c}$ limit, is a renormalizable Lagrangian.

Notice that, if the heavy constituent quark fields are integrated out  with the value of  $g_A$ left as a free parameter~\cite{PdeR93}, the resulting $O(p^4)$ couplings of the chiral Lagrangian, i.e. the $L_{i}$ of the Gasser--Leutwyler Lagrangian~\cite{GL85}, all become UV--divergent~\footnote{See ref.~\cite{BBdeR93} for a detailed discussion} and, therefore, the predictive power of the effective Lagrangian in Eq.~\rf{CCQL}, seen as a renormalizable Lagrangian, becomes rather restricted. Weinberg in his recent paper~\cite{Wei10} has only considered the chiral--${\rm SU(2)}$ case {\it without external sources} and this  is why with $g_A$ left as a free parameter he finds that  only two counterterms are needed in that case. With the choice $g_A =1$ these two terms have finite couplings.

\section{\normalsize Predictions of the Effective Lagrangian and the Value of  $M_Q$}
\setcounter{equation}{0}
\def\theequation{\arabic{section}.\arabic{equation}}

We wish to discuss some of the predictions of the Lagrangian in Eq.~\rf{CCQL}.
Integrating out  the heavy constituent quark fields  results in an effective Lagrangian where only the Goldstone modes are active. If the C$\chi$QM, or its version in Eq.~\rf{CCQL} with $g_{A}=1$, is a good effective Lagrangian of QCD, the resulting chiral Lagrangian of Goldstone modes alone should reproduce, in particular, the phenomenological determination of the $\cO (p^4)$  Gasser-Leutwyler low energy couplings. This integration was done in ref.~\cite{EdeRT90} and the fact is that, with $g_A=1$, the resulting  constants which are leading in the $1/{\rm N_c}$--expansion and  do not involve external scalar and/or pseudoscalar sources,   turn out to be  {\it finite}  and reproduce rather well the phenomenological values of the $L_{i}$--constants:
\be\lbl{conv}
2L_1 =L_2  =\frac{1}{12}\frac{N_c}{16\pi^2}\,,\quad L_3=\frac{1}{6}\frac{N_c}{16\pi^2}
\quad\annd\quad  2L_{10} =-L_9 =-\frac{1}{3}\frac{N_c}{16\pi^2}\,.
\ee
Had we used a value of $g_{A}\not= 1$, these constants will all be dominated by  logarithmically divergent terms proportional to $g_{A}-1$ and, hence, within Weinberg's  framework of a renormalizable Lagrangian~\cite{Wei10}, their corresponding couplings would have to be added as explicit counterterms, with a corresponding lost of predictive power.

Notice that the results in Eq.~\rf{conv} do not depend on the value of the constituent quark mass $M_Q$. The dependence on $M_Q$ appears  at $\cO(p^6)$ or higher in the chiral expansion. {\it What is the value of $M_Q$ that should be used in phenomenological applications?}
The underlying physical picture seems to us as follows:
because of confinement, the Lagrangian in Eq.~\rf{CCQL} cannot be trusted to evaluate QCD Green's functions in regions where $Q\bar{Q}$ pairs can be produced as free states. On the other hand, integrals of  $Q\bar{Q}$ pairs in the Minkowski region are related via dispersion relations to values of Green's functions in the Euclideann region where, at low momenta, the effective theory is expected to predict reasonable results, like e.g. the $L_{i}$--constants above.
Increasing $M_Q$ is equivalent to
 pushing the $Q\bar{Q}$  threshold in the Minkowski region to higher and higher values, which means increasing the mass gap between the massless Goldstones and the underlying hadronic spectrum. The mass gap in the hadronic world is provided by the
$\rho$-mass which is  $\sim 800~\MeV$. This suggests  $2 M_Q < M_{\rho}$ as an upper bound for $M_Q$. However, in the dispersion relation obeyed by the effective field theory, the area provided by the ``unconfined'' shape of the $Q\bar{Q}$ pairs in the Minkowski region should match approximatively  the phenomenological one provided by the  $\rho$--narrow state. In order to guarantee this matching one then has to lower the $Q\bar{Q}$ threshold with respect to $M_{\rho}$. Let us discuss this duality argument more quantitatively.

We suggest considering  the hadronic vacuum polarization due to the electromagnetic interactions of the light quarks
\be
i\int d^4 x e^{-ik\cdot x} \langle 0\mid T \left( J_{\mu}^{~\rm em}(x) J_{\nu}^{~\rm em}(0) \right) \mid 0\rangle = (k_{\mu} k_{\nu}-g_{\mu\nu}k^2 ) \Pi^{({\rm H})}(k^2)\,,
\ee
where  $J_{\mu}^{~\rm em}=\frac{2}{3}{\bar u}\gamma_{\mu}u-\frac{1}{3}{\bar d}\gamma_{\mu}d-\frac{1}{3}{\bar s}\gamma_{\mu}s$. More precisely,  let us consider the slope at the origin of the  correlation function $\Pi^{({\rm H})}(k^2)$ i.e.
\be
\frac{\partial \Pi^{({\rm H})}(k^2)}{\partial k^2}\Big\vert_{k^2 =0} = \int_{0}^{\infty}\frac{dt}{t^2}\ \frac{1}{\pi}\Imm\Pi^{({\rm H})} \left(t\right)\,.
\ee
which is an $\cO(p^6)$ observable in the chiral expansion.
We then ask that the phenomenological value for the slope provided by the narrow width $\rho$--dominance approximation  to the hadronic spectral function~\footnote{ The overall factor $2/3$ comes from modulating the electric charge squared by the sum   of the $u$, $d$ and $s$  quark charges and $f_{\rho}$ is the $\rho$ to vacuum coupling constant related to the electronic width of the $\rho$--meson:
$
\Gamma_{\rho\ra e^+ e^-}=\frac{4\pi\alpha^2}{3}f_{\rho}^2 M_{\rho}\,.
$.}~:
\be
\frac{1}{\pi}\Imm\Pi^{({\rm H})} \left(t\right)\simeq\frac{2}{3}e^2\ 2f_{\rho}^2 M_{\rho}^2 \delta(t-M_{\rho}^2)\,,
\ee
matches the slope predicted by the C$\chi$QM.
This fixes $M_Q$ as follows
\be
M_Q^2 \simeq \frac{N_c}{15}\frac{1}{8\pi^2 f_{\rho}^2}M_{\rho}^2\,,
\ee
which for the observed values of $f_{\rho}$ and $M_{\rho}$, with their errors~\cite{PPB}, results in a rather low mass: 
\be
M_Q \simeq (194\pm 24)~\MeV\,.
\ee
We then suggest to take this estimate as the value of $M_{Q}$ , to which we add in quadrature a typical Large--$N_c$ error $\sim \frac{\Gamma_{\rho}}{M_{\rho}}= 20\%$ and round numbers to:
\be
M_{Q}=(190\pm 40)~\MeV\,.
\ee

Let us now discuss other predictions of the C$\chi$QM and see if they can be digested with this range of values for $M_Q$  when confronted to the phenomenological determinations.

\section{\normalsize The Left--Right Correlation Function}
\setcounter{equation}{0}
\def\theequation{\arabic{section}.\arabic{equation}}

This is the Green's function
\be\lbl{PiLR}
\Pi_{\rm LR}^{\mu\nu}(q)=2i\int d^4 x\,e^{iq\cdot x}\langle 0\mid
T\left(L^{\mu}(x)R^{\nu}(0)^{\dagger}\right)\mid 0\rangle 
\ee
of  left and right currents:
\be
L^{\mu}(x)=\bar{d}(x)\gamma^{\mu}\frac{1}{2}(1-\gamma_{5})u(x)
\qquad \annd \qquad
R^{\mu}(x)=\bar{d}(x)\gamma^{\mu}\frac{1}{2}(1+\gamma_{5})u(x)\,.
\ee
In the chiral limit where the light quark masses are set to zero ($Q^2=-q^{2}\ge 0$
for
$q^2$ spacelike)
\be\lbl{pi}
\Pi_{\rm LR}^{\mu\nu}(q)=(q^{\mu}q^{\nu}-g^{\mu\nu}q^2)\Pi_{\rm LR}(Q^2)\,,
\ee
and the self--energy function
$\Pi_{\rm LR}(Q^2)$ vanishes order by order in perturbation theory; it becomes  an
order parameter of the spontaneous breakdown of chiral symmetry for all
values of the momentum transfer~\cite{FSS90,KdeR98}. 

The Taylor expansion   of $\Pi_{\rm LR}(Q^2)$  at low $Q^2$ values is 
governed by successive couplings of the effective chiral Lagrangian of QCD:
\be
-Q^2\Pi_{\rm LR}(Q^2)\underset{{Q^2\ \ra 0}}{\thicksim}\ F_{\pi}^2+4L_{10} Q^2+8 C_{87} Q^4+\cO(Q^6)\,,
\ee
where $F_{\pi}$ is the pion coupling constant (to be identified with the $F_{\pi}$ which appears in the second line  of Eq.~\rf{CCQL}) and   
$L_{10}$ is the coupling which, for $g_A =1$, the C$\chi$QM predicts the value in Eq.~\rf{conv} i.e.
\be
L_{10}=-\frac{1}{6}\frac{N_c}{16\pi^2}=-3.2\times 10^{-3}\,,
\ee
to be compared with the phenomenological determination~\cite{GPP10}
\be
L_{10}^{r}(M_{\rho})=-(4.05\pm 0.39)\times 10^{-3}\,,
\ee
of the renormalized coupling at the $\rho$--mass scale.

 The constant $C_{87}$ corresponds to a  specific coupling  of  $\cO(p^6)$~\cite{ABT00}. It has also been determined phenomenologically, using data from hadronic $\tau$ decays, with the result~\cite{GPP10}
\be
C_{87}(M_{\rho})=(4.88\pm 0.13)\times 10^{-3}~\GeV^{-2}\,.
\ee
The C$\chi$QM prediction for $M_Q =(190\pm 40)~\MeV$ is 
\be
C_{87}=\frac{N_c}{16\pi^2}\frac{1}{120 M_{Q}^2}=(4.4\pm 1.8)\times 10^{-3}~\GeV^{-2}\,.
\ee
We, therefore, conclude that the C$\chi$QM predictions for $L_{10}$ and $C_{87}$, considered as a first estimate, are rather good . 

In principle, due to the fact that $\Pi_{\rm LR}(Q^2)$ is an order parameter of S$\chi$SB, one expects the predictions of the C$\chi$QM to the low--$Q^2$ behaviour of this function to become worse and worse as the power of the $Q^2$ series, corresponding to higher and higher couplings in the chiral expansion, increases. Related to that is the presence of the coupling $e^2 { C}\ \tr (Q_R U Q_L U^{\dagger})$ in Eq.~\rf{CCQL}. This is the coupling which gives a mass of electromagnetic origin to the charged pion in the chiral limit, and it is fixed by the integral~\cite{Lowetal67,KPdeR..} 
\be\lbl{piem}
m_{\pi^+}^{2}\vert_{\footnotesize\rm em}=-\frac{2e^2 C}{F_{\pi}^2}=
\frac{\alpha}{\pi}\,\frac{3}{8F_{\pi}^2}\,
\int_0^\infty
dQ^2\,\left[-Q^2\Pi_{\rm LR}(Q^2)\right]\ .
\ee
In QCD this integral converges in the ultraviolet because~\cite{SVZ79}
\be\lbl{OPE}
\lim_{Q^2\ra\infty}\Pi_{\rm LR}(Q^2)\sim \cO\left(\frac{\stern^2}{Q^6}\right)\,,
\ee
while in the C$\chi$QM

{\setl 
\bea
-Q^2 \Pi_{\rm LR}(Q^2)\big\vert_{{\rm C}\chi {\rm QM}} & = &
 F_{\pi}^2-\frac{N_c}{4\pi^2}M_{Q}^2
\int_{0}^\infty dx \log\left[1+\frac{Q^2}{M_{Q}^2}x(1-x) \right]\nn \\
& = &
 F_{\pi}^2+\frac{N_c}{4\pi^2}M_{Q}^2 \left(2+
\sqrt{1+\frac{4M_{Q}^2}{Q^2}}\log\frac{\sqrt{1+\frac{4M_{Q}^2}{Q^2}}-1}{\sqrt{1+\frac{4M_{Q}^2}{Q^2}}+1} \right)\,;
\eea}

\noi 
and for large $Q^2$
\be
-Q^2 \Pi_{\rm LR}(Q^2)\big\vert_{{\rm C}\chi {\rm QM}}  \underset{{Q^2\ \ra\infty}}{\thicksim}\ F_{\pi}^2 -\frac{N_c}{4\pi^2}M_{Q}^2 \left(\log\frac{Q^2}{M_Q^2}-2 \right)+\cO\left[\frac{M_{Q}^4}{Q^2}\log\frac{Q^2}{M_Q^2} \right]
\ee
which fails, dramatically, to match the QCD short--distance behaviour. As a result the integral in Eq.~\rf{piem} diverges quadratically in the C$\chi$QM.  This is why one needs the explicit local coupling $e^2 { C}\ \tr (Q_R U Q_L U^{\dagger})$ in the low energy effective Lagrangian with a coupling $C$ which, like $F_{\pi}$ and $B$, has to be fixed phenomenologically. 

There is in fact another observation one can make from this result. It has to do with the fact that in QCD~\cite{W83, CLT95} 
\be\lbl{witten}
-Q^2\Pi_{\rm LR}(Q^2)\ge 0 \qquad\foor\;  \quad 0\le Q^2\le\infty\,,
\ee
which in particular ensures the positivity of the integral in
Eq.~\rf{piem} and thus the stability of the QCD vacuum with respect
to  small perturbations induced by electromagnetic interactions. The C$\chi$QM, however, does not satisfy this positivity constraint since eventually, at asymptotically large--$Q^2$ values, $-Q^2 \Pi_{\rm LR}(Q^2)\big\vert_{{\rm C}\chi {\rm QM}}$ becomes negative. In fact for $M_Q \simeq 200~\MeV$ this already occurs at $Q\simeq 2~\GeV$.  This failure of the C$\chi$QM is quite a generic one; it is likely to happen whenever one has to deal with integrals over the whole euclidean range of Green's functions which are order parameters of S$\chi$SB. This is why  it is difficult to attribute much significance  to calculations of {\it most} of the couplings of the electroweak hadronic Lagrangian which, in the literature, have been made within the framework of  constituent chiral quark inspired models. 

The {\it most} in the previous paragraph means that there are, however, {\it exceptional} cases, as already mentioned in the introduction. The next section is dedicated to one such case. Other {\it exceptional} cases are the  contributions to $g_{\mu}-2$ from hadronic vacuum polarization, from the $Z\gamma\gamma$--triangle, and from hadronic light--by--light scattering. These other contributions have also been calculated in the C$\chi$QM and will be discussed in  detail in a forthcoming paper~\cite{GdeR01}.

\section{\normalsize The Decay $\pi^0 \ra e^+ e^-$ in the C$\chi$QM.}
\setcounter{equation}{0}
\def\theequation{\arabic{section}.\arabic{equation}}

The discussion of this process from the point of view of the low energy effective chiral Lagrangian of QCD, can be found in ref.~\cite{KPPdeR99}. The transition amplitude is dominated by the exchange of two intermediate photons and, hence, it is convenient to consider its decay rate normalized to the one of the $\pi^0 \ra \gamma\gamma$ decay rate (with $N_c =3$):
\be\lbl{BR}
{\rm Br}\equiv
\frac{\Gamma\left(\pi^0 \ra e^+ e^-\right)}{\Gamma\left(\pi^0 \ra \gamma\gamma\right)} = 2\left(\frac{\alpha}{\pi} \right)^2 \frac{m_e^2}{M_{\pi}^2}\beta(M_{\pi}^2)\mid\cA(M_{\pi}^2)\mid^2\,,
\ee
where $\beta(s)=\sqrt{1-\frac{4m_{e}^2}{s}}$ and $\cA(M_{\pi}^2)$ a dynamical amplitude which, to lowest order in the chiral expansion, has the form~\cite{KPPdeR99}:
\be\lbl{ampl}
\cA(s)=\frac{N_c}{3}\left[\frac{3}{2}\log\left(\frac{m_{e}^2}{\mu^2}\right)-\frac{5}{2}+\cC(s)  \right]+\chi(\mu)\,.
\ee
Here the function $\cC(s)$ results from the calculation of the triangle loop  with the Adler, Bell-Jackiw  point--like $\pi^0 \gamma\gamma$ coupling. This loop is divergent and the $\mu$--scale in the $\log$--term is the one associated  to the loop amplitude renormalized  in the ${\overline{\rm MS}}$--scheme of dimensional regularization. Then~\footnote{The expression for $s>0$ is the one which follows from analytic continuation with the usual $i\epsilon$--prescription.}, for $s<0$ and $\Li_2 (x)=-\int_0^{x} (dt/t)\log(1-t)$,
\be 
\cC(s)=\frac{1}{\beta(s)}\left[\Li_2 \left(\frac{\beta(s)-1}{\beta(s)+1}\right)
+\frac{\pi^2}{3}+\frac{1}{4}\log^2 \left(\frac{\beta(s)-1}{\beta(s)+1}\right)\right]\,.
\ee
The coupling $\chi(\mu)$ is a renormalized coupling constant related to the two couplings $\chi_1$ and $\chi_2$ which, to lowest order in the chiral expansion, describe the direct local interactions~\cite{SLW92} of Goldstone fields with leptons ($l=e,\mu$):

{\setl
\bea
\lefteqn{\hspace*{3cm}\cL_{U l^+ l^- }(x)=i\frac{3}{32}\left(\frac{\alpha}{\pi} \right)^2 \bar{l}(x)\gamma^{\mu}\gamma_{5}l(x)\times} \nn\\ 
& & \hspace*{-0.5cm}\left[\chi_1 \tr \left(Q_R Q_R D_{\mu}U\ U^{\dagger}\!-\!Q_L Q_L  D_{\mu}U^{\dagger}\ U\right)\! +\!\chi_{2} \tr\left(U^{\dagger}Q_R D_{\mu}UQ_L \!-\! U Q_{L}D_{\mu}U^{\dagger}Q_R \right)\right]\,.
\eea}

\noi
In fact: $\chi=-\frac{1}{4}(\chi_1 +\chi_2 )$ and
 the scale dependence of the renormalized coupling $\chi(\mu)$, in the same ${\overline{\rm MS}}$--scheme of dimensional regularization, cancels with the one in the $\log$--term in Eq.~\rf{ampl}. 

\begin{figure}[h]

\begin{center}
\includegraphics[width=0.5\textwidth]{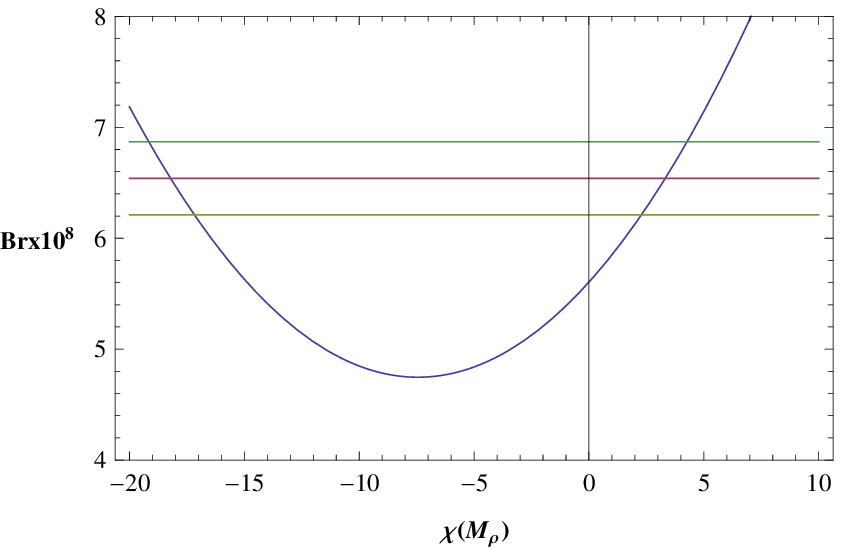}

\vspace*{0.25cm}
{\bf Fig.~1}
{\it\small  Plot of the Branching Ratio  in Eq.~\rf{BR} versus $\chi(M_{\rho})$.
}
\end{center}
\end{figure}
\noi
The predicted branching ratio in Eq.~\rf{BR} as a function of $\chi$ at the scale of the $\rho$--mass is shown in Fig.~1.
The experimental value of Br~\cite{PPB}:
\be\lbl{brexp}
{\rm Br}=(6.54\pm 0.33)\times 10^{-8}\,,
\ee
is also shown in Fig.~1 (the horizontal strip).
This fixes a twofold solution for $\chi(M_{\rho})$, one negative the other positive. Let us now discuss what theory predicts for this $\chi$--coupling.

The underlying Green's function of this process is the correlation  function~\cite{KPPdeR99}:

{\setl
\bea
\lefteqn{\hspace*{-1cm}
\int d^4 x \!\!\int d^4 y e^{iq_1 \cdot x} e^{iq_2 \cdot y}\langle 0\mid T\large( J_{\mu}^{~\rm em}(x)J_{\nu}^{~\rm em}(y)P^{(3)} (0)\large)\mid 0\rangle} \nn \\ & & =
\epsilon_{\mu\nu\alpha\beta}q_1 ^\alpha q_2 ^\beta 
i\frac{\stern}{F_{\pi}^2}\frac{i}{(q_1  +q_2)^2 }\cF_{\pi^0 \gamma*\gamma*}[(q_1 +q_2)^2,q_1^2 ,q_2^2]\,,
\eea}

\noi
where $P^{(3)}=(\bar{u}i\gamma_{5}u-\bar{d}i\gamma_{5}d)/2$ and $\cF_{\pi^0 \gamma*\gamma*}[(q_1 +q_2)^2,q_1^2 ,q_2^2]=\cF_{\pi^0 \gamma*\gamma*}[(q_1 +q_2)^2,q_2^2 ,q_1^2]$ is the off--shell $\pi^0 \gamma\gamma$ form factor. The coupling $\chi(\mu)$ is then given by the residue of the pole of the matrix element
\be
 \langle e(p')\mid P^{(3)}(0)\mid e(p)\rangle\quad {\rm at}\quad  (p'-p)^2 =0\,,
\ee
after subtraction of the two on--shell photon loop. Keeping only the contributions that do not vanish as $(p' -p)^2 \ra 0$ and neglecting terms of $\cO(m_{l}^2 /\Lambda_{\rm H}^2)$ where $\Lambda_{\rm H}$ denotes a scale of hadronic origin, one gets

{\setl
\bea\lefteqn{
\frac{\chi(\mu)}{32\pi^4}   =   -4\ i \left( 1-\frac{1}{d}\right)  \lim_{(p'-p)^2 \ra 0}
\int \frac{d^d k}{(2\pi)^d}\left(\frac{1}{k^2} \right)^2} \nn \\
 & & \times\left\{\cF_{\pi^0 \gamma*\gamma*}[(p' -p)^2,k^2 ,(p'-p-k)^2]-\cF_{\pi^0 \gamma*\gamma*}[(p' -p)^2,0 ,0]\right\} \lbl{chiexp} \,.
\eea}

If now one assumes that the underlying dynamics is  governed by the C$\chi$QM we simply have to introduce the expression
\be\lbl{FFCQ}
\cF_{\pi^0 \gamma*\gamma*}^{{\rm C}\chi{\rm QM}}(0,k^2 ,k^2)
=-\frac{N_c}{12\pi^2}\int_0^1 dz \frac{M_Q^2}{M_Q^2 -z(1-z)k^2}
\ee
in Eq.~\rf{chiexp}. This expression has the correct QCD normalization  at the origin; however, in the deep euclidean region it behaves as 
\be
\cF_{\pi^0 \gamma*\gamma*}^{{\rm C}\chi{\rm QM}}(0,k^2 ,k^2) \underset{{-k^2\ \ra\infty}}{\thicksim}\
-\frac{N_c}{12\pi^2}2\frac{M_Q ^2}{-k^2}\log\left(\frac{-k^2}{M_Q ^2}\right)
\ee 
while, as shown in ref.~\cite{KPPdeR99}, the asymptotic behaviour predicted by the  OPE, is
\be
\cF_{\pi^0 \gamma*\gamma*}^{{\rm OPE}}(0,k^2 ,k^2) \underset{{-k^2\ \ra\infty}}{\thicksim}\
-\frac{2}{3}\frac{F_{\pi}^2}{-k^2}\,;
\ee 
however, unlike the case of the Left--Right Correlation Function, this mismatch is not too bad. Proceeding ahead with the C$\chi$QM form factor in Eq.~\rf{FFCQ}
results in
\be
\chi^{\rm C\chi\rm QM}(\mu)=\frac{N_c}{2}\left(\log \frac{\mu^2}{M_Q^2}-\frac{7}{6}\right)\,;
\ee
and therefore, with $M_Q =(190\pm40)~\MeV$ and  $\mu=M_{\rho}$,  which is the appropriate choice of scale so as to compare with previous estimates, we get
\be\lbl{chiqcm}
\chi^{\rm C\chi\rm QM}(M_{\rho})=2.5\pm 0.6\,,
\ee
remarkably  close to the {\it lowest meson dominance} approximation to Large--$N_c$ estimate made in ref.~\cite{KPPdeR99}: $\chi^{\rm LMD}(M_{\rho})=2.2\pm 0.9$.

\begin{figure}[h]

\begin{center}
\includegraphics[width=0.5\textwidth]{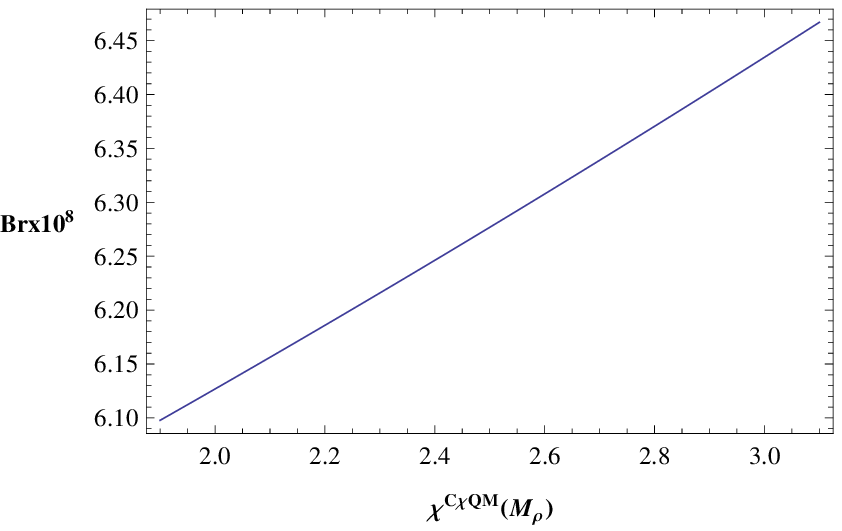}

\vspace*{0.25cm}
{\bf Fig.~2}
{\it\small  Predicted Br in Eq.~\rf{BR}  versus the value of $\chi^{\rm C\chi\rm QM}(M_{\rho})$ in Eq.~\rf{chiqcm}.
}
\end{center}
\end{figure}

\noi
Figure 2 shows the branching ratio in Eq.~\rf{BR} predicted by the C$\chi$QM value of $\chi(M_{\rho})$ in the range corresponding to Eq~\rf{chiqcm}. The predicted Br is to be compared with the experimental value in Eq.~\rf{brexp}.
We conclude that the C$\chi$QM does very well in digesting the  $\pi^0 \ra e^+ e^-$ decay rate, with a constituent quark mass in the range: $M_Q =(190\pm 40)~\MeV$.

\section{\normalsize Conclusions.}
\setcounter{equation}{0}
\def\theequation{\arabic{section}.\arabic{equation}}
From the previous considerations we conclude that the simple C$\chi$QM Lagrangian of Eq.~\rf{CCQL}, with $g_{A}=1$, does rather well as an effective Lagrangian of Large--${\rm N_c}$ QCD at low energies. As emphasized by Weinberg, it has the nice feature that, to leading order in the Large--${\rm N_c}$ limit, it is a renormalizable Lagrangian and, with $g_{A}=1$, only a few counterterms are needed. The predicted values of the five $\cO(p^4)$ Gasser--Leutwyler couplings, which for $g_A =1$ are finite, are within the bulk of the phenomenological determinations.  

We  have presented a duality argument to fix the constituent quark mass and found a value in the range $M_Q =(190\pm 40)~\MeV$. With this determination we find that the C$\chi$QM prediction for the $\cO(p^6)$ coupling $C_{87}$ reproduces rather well the phenomenological determination from hadronic $\tau$--decays. 

We have also discussed the limitations of the model as well as the {\it exceptional cases} of applications to low--energy observables involving the integration of Green's functions over the full range of euclidean momenta, where the model can still be expected to produce reasonable predictions. The decay $\pi^0 \ra e^+ e^-$ is one such  example which we have discussed in detail.

\vspace*{0.5cm}

\begin{center}
{\normalsize\bf Acknowledgements.}
\end{center}

I wish to thank M.~Knecht and S.~Peris for many useful discussions. This work has been partially supported by the EU RTN network FLAVIAnet [Contract No. MRTN-CT-2006-035482].


\vfill

\end{document}